# Coevolving Boolean and Multi-Valued Regulatory Networks

Larry Bull

Computer Science Research Centre

University of the West of England, Bristol UK

Larry.Bull@uwe.ac.uk

**Abstract**

Random Boolean networks have been used widely to explore aspects of gene regulatory networks. A modified form of the model through which to systematically explore the effects of increasing the number of gene states has previously been introduced. In this paper, these discrete dynamical networks are coevolved within coupled, rugged fitness landscapes to explore their behaviour. Results suggest the general properties of the Boolean model remain with higher valued logic regardless of the update scheme or fitness sampling method. Introducing topological asymmetry in the coevolving networks is seen to alter behaviour.

## 1. Introduction

Coevolution is ubiquitous in natural systems and the well-known abstract NKCS model [Kauffman & Johnsen, 1992] has long been used to systematically explore some of the basic properties of such systems. Extending the NK model [Kauffman and Levin, 1987] of rugged fitness landscapes, the NKCS model couples multiple NK landscapes to capture the evolutionary dynamics of ecosystems containing multiple species. Random Boolean networks (RBN)[Kauffman, 1969] have been previously been used in conjunction with both the NK and NKCS models to explore the evolution of gene regulation within tuneable fitness landscapes [Bull, 2012]. The standard NK and NKCS models assume a binary gene alphabet but a recent extension to higher alphabets suggests the basic properties of the original models remain [Bull, 2022]. The effects of altering the size of the alphabet of the underlying gene expression state representation and logic in RBN has recently been explored using the non-binary NK model [Bull, 2023] (after [Solé et al., 2000]). Results suggest that a number of the basic properties of the original binary model remain, whilst aspects such as how fitness is sampled and how many genes contribute explicitly to the fitness calculation can significantly vary behaviour. This paper explores the coevolution of multi-valued networks finding the basic properties remain, although there is an increase in sensitivity to the degree of internal network connectivity. Asymmetric properties in the partners are also explored, with mutually beneficial degrees of network connectivity and effects due to relative regulatory network size found.

## 2. Random Multi-Valued Networks

Within the traditional form of RBN, a network of $R$ nodes, each with $B$ directed connections randomly assigned from other nodes in the network, all update synchronously based upon the current state of those *B* nodes. As the name suggests, gene states are traditionally from a binary alphabet (*A*=2) and use a randomly assigned Boolean update function. Hence those *B* nodes are seen to have a regulatory effect upon the given node, specified by the Boolean function attributed to it. Since they have a finite number of possible states and they are deterministic, such networks eventually fall into an attractor. It is well-established that the value of *B* affects the emergent behaviour of RBN wherein attractors typically contain an increasing number of states with increasing *B* (see [Kauffman, 1993] for an overview). Three regimes of behaviour exist: ordered when *B*=1, with attractors consisting of one or a few states; chaotic when *B*>2, with a very large number of states per attractor; and, a critical regime around *B*=2, where similar states lie on trajectories that tend to neither diverge nor converge (see [Derrida & Pomeau, 1986] for formal analysis). Note that the size of an RBN is traditionally labelled *N*, as opposed to *R* here, and the degree of node connectivity labelled *K*, as opposed to *B* here. The change is adopted due to the traditional use of the labels *N* and *K* in the NKCS model of fitness landscapes which are also used in this paper, as will be shown.

As noted above, multi-valued logic forms of the original binary model have been explored. Following [Bull, 2023], in the simplest case, each node can exist in one of *A* (*A*≥2) states and is assigned a randomly created logic table for each of the $A^B$ possible configurations (Figure 1). Figure 2 shows the typical number of nodes changing state per update cycle in such discrete dynamical systems where *R*=50, with various connectivity *B* and number of gene expression states *A*, using 0<*B*<6 and 1<*A*<9. As can be seen, in these random multi-valued networks (RMN), for high connectivity (*B*>2) behaviour is significantly changed with increasing *A*. That is, significantly more nodes change state per update cycle when *A*>2 with such connectivity. Formal analysis suggests the critical regime occurs at *B*=1 with increasing *A* [Solé et al., 2000].

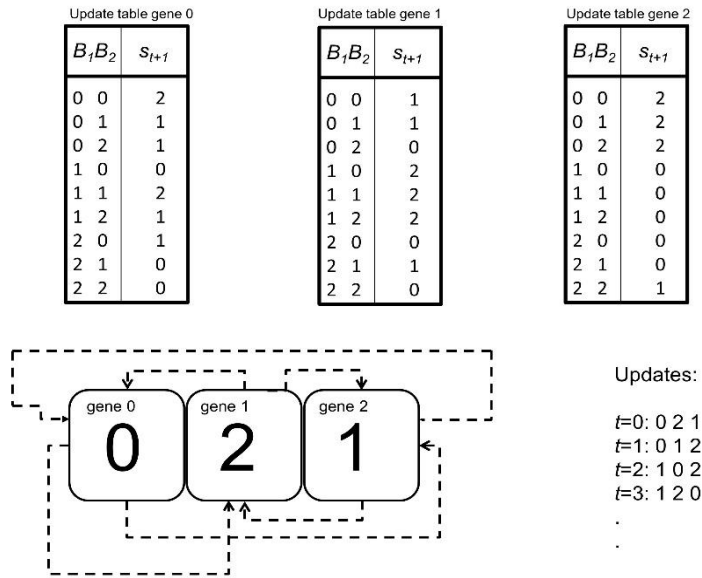

Figure 1. An example random multi-valued regulatory network model, with

$R$=3, $B$=2 and $A$=3.

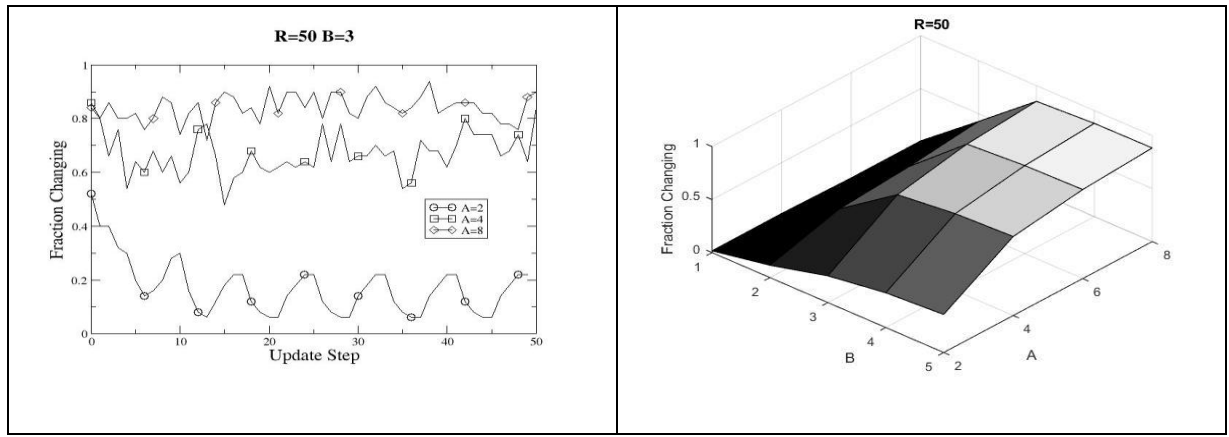

Figure 2. Showing the effects on the typical behaviour of the multi-valued regulatory networks with varying connectivity $B$ and states $A$. Results are the average of one hundred randomly created networks per parameter configuration.

## 3. The NKCS Model

Kauffman and Levin [1987] introduced the NK model to allow the systematic study of various aspects of fitness landscapes. In the standard model, the features of the fitness landscapes are specified by two parameters: $N$, the length of the genome; and $K$, the number of genes that has an effect on the fitness contribution of each binary gene ($A=2$). Thus increasing $K$ with respect to $N$ increases the epistatic linkage, increasing the ruggedness of the fitness landscape. The increase in epistasis increases the number of optima, increases the steepness of their sides, and decreases their correlation [Kauffman, 1993]. The model assumes all intragenome interactions are so complex that it is only appropriate to assign random values to their effects on fitness. Therefore for each of the possible $K$ interactions a table of $A^{(K+1)}$ fitnesses is created for each gene with all entries in the range 0.0 to 1.0, such that there is one fitness for each combination of traits. The fitness contribution of each gene is found from its table. These fitnesses are then summed and normalized by $N$ to give the selective fitness of the total genome.

Kauffman and Johnsen [1992] subsequently introduced the abstract NKCS model to enable the study of various aspects of *co*evolution. Each gene is said to also depend upon $C$ randomly chosen traits in each of the other $S$ species with which it interacts. Altering $C$, with respect to $N$, changes how dramatically adaptive moves by each species deform the landscape(s) of its partner(s), where increasing $C$ typically increases the time to equilibrium. Again, for each of the possible $K+(S \times C)$ interactions, a table of $A^{(K+(S \times C)+1)}$ fitnesses is created for each gene, with all entries in the range 0.0 to 1.0, such that there is one fitness for each combination of traits. Such tables are created for each species (Figure 3).

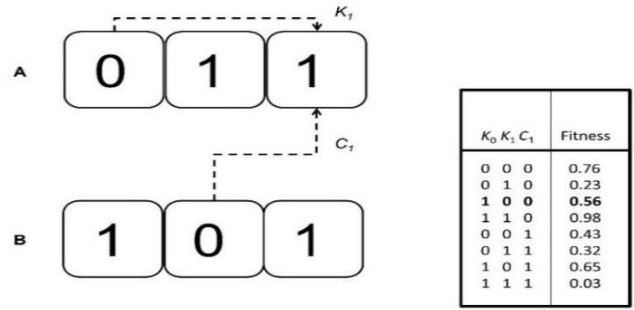

Figure 3. The traditional NKCS model: Each gene is connected to *K* randomly chosen local genes and to *C* randomly chosen genes in each of the *S* other species. Connections and table shown for one gene in one species for clarity. Here *N*=3, *K*=1, *C*=1, *S*=1, *A*=2.

Following [Kauffman, 1993], a mutation-based hill-climbing algorithm is used here, where the single point in the fitness space is said to represent a converged species, to examine the properties and evolutionary dynamics of the NKCS models. That is, the population is of size one and a species evolves by making a random change to a randomly chosen genes per generation. The "population" is said to move to the genetic configuration of the mutated individual if its fitness is greater than the fitness of the current individual; the rate of supply of mutants is seen as slow compared to the actions of selection. Ties are broken at random.

Figure 4 shows example behaviour for one of two coevolving species where the parameters of each are the same and hence behaviour is symmetrical. The effects of mutual fitness landscape movement are clearly seen. All results reported in this paper are the average of 10 runs (random start points) on each of 10 randomly created NKCS fitness landscapes, i.e., 100 runs, for each parameter configuration. The average final fitness of the converged population is used for graphs. Here 0≤*K*≤10, 1≤*C*≤5, *S*=1, for *N*=20 and *N*=100. Figure 5 shows how increasing the degree of coupling (*C*) between the two landscapes causes fitness levels to fall significantly (T-test, *p*<0.05) when *C*≥*K* for *N*=20. Note this change in behaviour around *C*=*K* was suggested as significant in [Kauffman & Johnson, 1992], where *N*=24 was used throughout. However, Figure 5 also shows how with *N*=100 fitness *always* falls significantly with increasing *C* (T-test, *p*<0.05), regardless of *K*. As noted in [Kauffman & Johnson, 1992], increasing *N* increases the time to equilibrium; the duration of evolutionary search is increased. Figure 6 shows an example of how the same basic behaviour is seen when the alphabet is increased (*A*=4)(see [Bull, 2022] for further details).

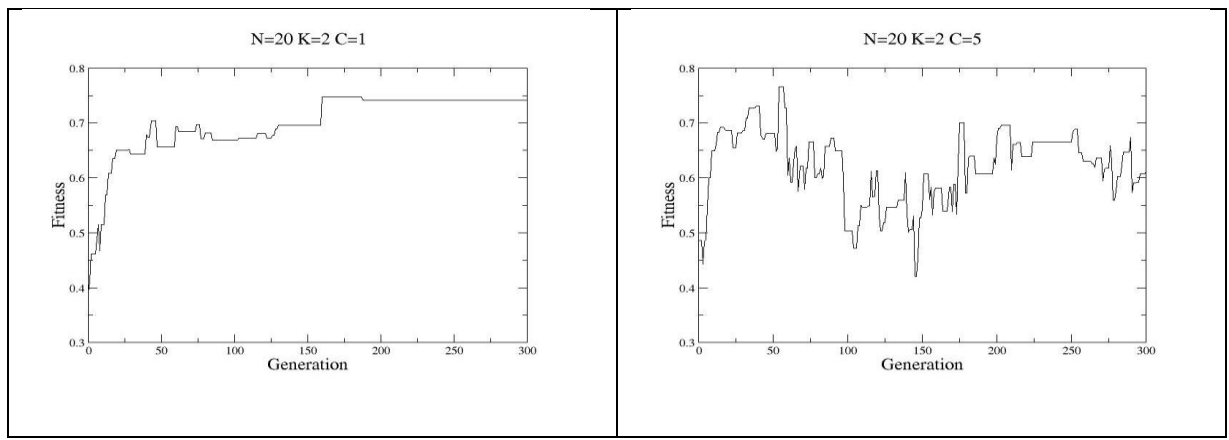

Figure 4. Showing example single runs of the typical behaviour of the standard NKCS model of coevolution with different degrees of landscape coupling (*C*) and *A*=2.

There is some previous work on the coevolving regulatory networks, beginning with Kauffman [1993] who briefly described the use of evolution to design RBN which must play a pair-wise (mis)matching game where mutation is used to change connectivity, the Boolean functions, *B*, and *R* (see [Bull, 2012] for a review). The NKCS model was previously used with RBN with different fitness and update schemes [Bull, 2012]. This paper expands that work to consider higher alphabets of states and asymmetry in the coupled networks.

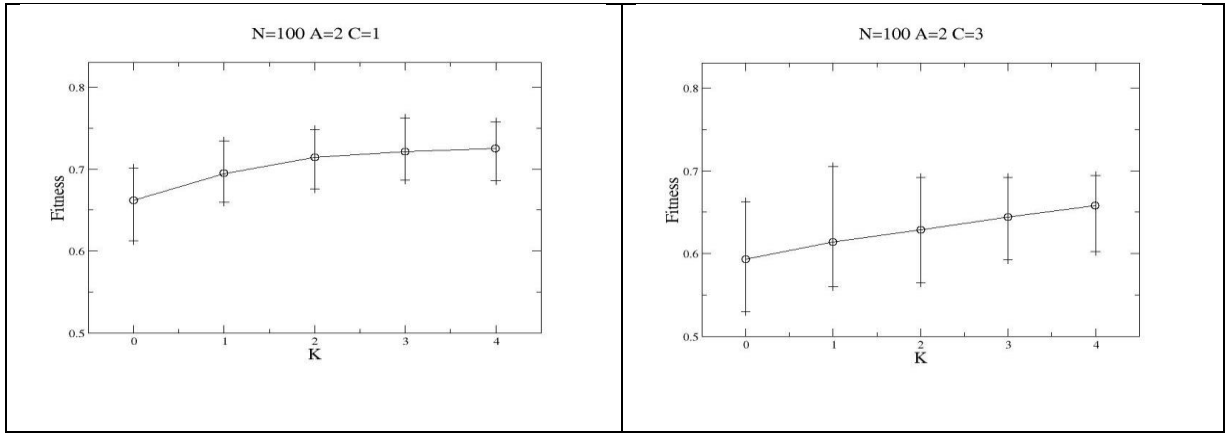

Figure 5. Showing the fitness reached after 20,000 generations on landscapes of varying ruggedness (*K*) and coupling (*C*). Here *A*=2. Error bars show min and max values.

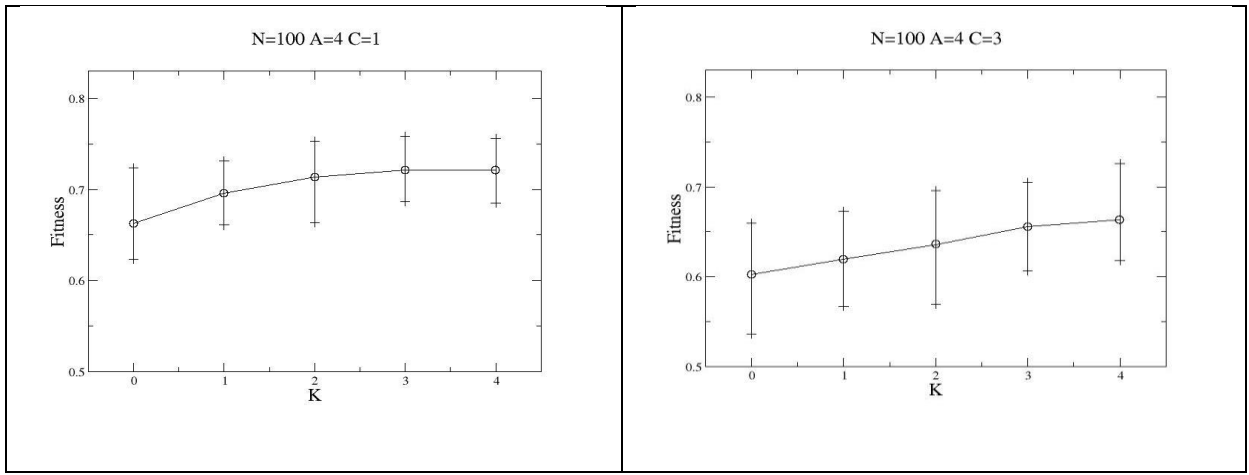

Figure 6: Showing the fitness reached after 20,000 generations on landscapes of varying ruggedness (*K*), coupling (*C*), and length (*N*), with a quaternary alphabet (*A*=4).

## 4. The RMNKCS Model

The combination of the discrete dynamical networks and the NKCS model enables the exploration of the relationship between phenotypic traits and the genetic regulatory network by which they are produced [Bull, 2012]. In this paper, the following simple scheme is adopted: *N* phenotypic traits are attributed to the first *N* nodes within the network of *R* genes (where 0<*N*≤*R*, Figure 7). Thereafter all aspects of the two models remain as described above, with simulated evolution used to evolve the RMN on NKCS landscapes. Hence the NKCS element creates an explicitly tuneable component to the RMNs' fitness landscapes and the activity of the GRN primarily affect, and are affected by, other GRN.

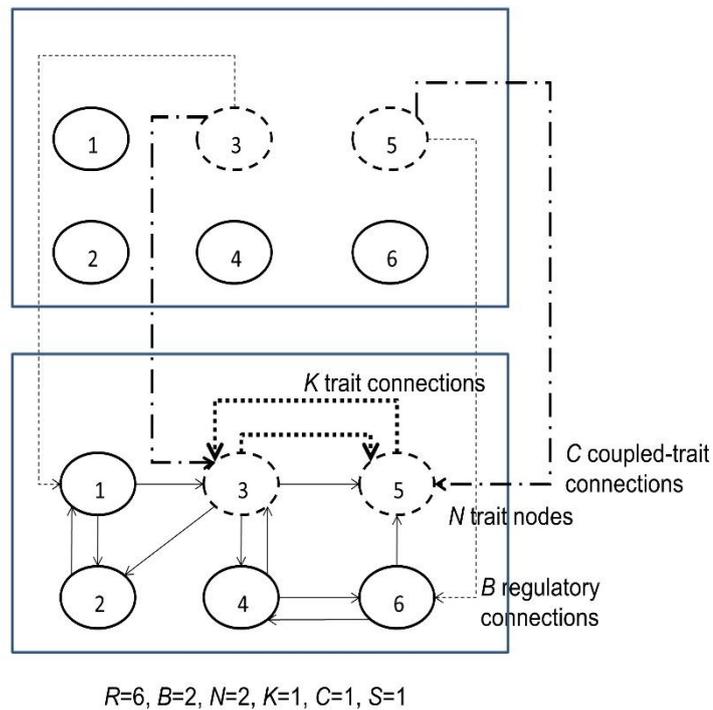

*R*=6, *B*=2, *N*=2, *K*=1, *C*=1, *S*=1

Figure 7. Example *RMNKCS* model. Connections for only one of the two coupled networks are shown for clarity.

For a given RMN, mutation can either alter the logic function of a randomly chosen node or alter a randomly chosen connection for that node (equal probability). A single fitness evaluation of a set of RMN is ascertained by first assigning each network node to a randomly chosen start state (uniform in $A$) and updating each RMN in turn for $U$ cycles. Here $U$ is chosen such that the networks have typically reached an attractor ($U=50$). At update cycle $U$, the value of each of the $N$ trait nodes of each RMN is then used to calculate fitness on the given NKCS landscape. In turn, a mutated RMN is evaluated with the current other species/RMN and becomes the parent for that species for the next generation if its fitness is higher than that of the original (ties are broken at random, all coevolving species update their fitnesses if a mutant is adopted). This process is repeated ten times on the given NKCS landscape, repeated for ten randomly created NKCS landscapes, i.e., 10x10=100 runs, with the fitness assigned to the RMNs being the average fitness after 5000 generations.

Figure 8 shows the typical coevolutionary performance of one species of a symmetrical pair of $R=50$ RMN with various internal connectivity ($0<B<6$), landscape ruggedness ($0≤K<6$) and coupling ($0<C<6$), with alphabet $A=2$, after 5000 generations. When $N=10$ (left column), fitness generally decreases with increasing $B$, regardless of $C$, as somewhat predicted by the results in Figure 2. Conversely, fitness typically increases with increasing $K$, regardless of $C$. Hence the highest fitness levels are seen for low $B$ and high $K$, regardless of $C$. That is, results for $B=1$ or $B=2$ are always statistically better (T-test, $p<0.05$) than for $B=4$ or $B=5$, except when $C=5$ where it is only the case for $K>4$. Figure 8 also shows results for when all RMN nodes make an explicit contribution to the fitness calculation ($R=N$). The same general behaviour is seen as before, although fitnesses are significantly lower for $B=4$ or $B=5$ in comparison to when $N=10$ (T-test, $p<0.05$).

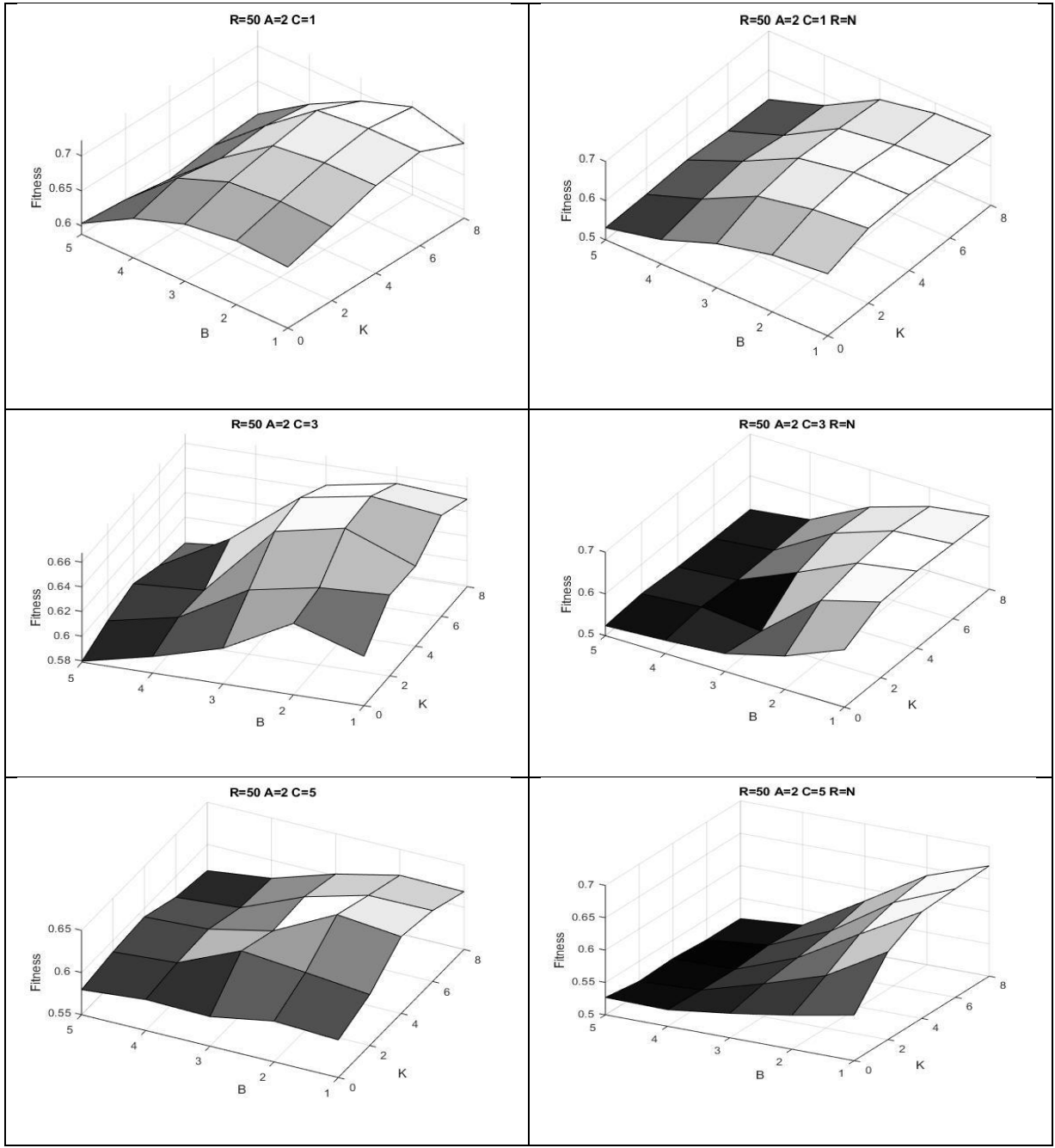

Figure 8. Showing the fitness reached by one species after 5000 generations on landscapes of varying ruggedness (*K*), coupling (*C*), number of explicit fitness traits (*N*), with *A*=2.

Figure 9 shows examples of how results are generally the same when increasing *A*. As previously reported in the non-coevolutionary case of a single, static fitness landscape [Bull, 2023], fitness typically significantly increases for *B*=1 with *A*>2 (T-test, *p*<0.05) when *R*=*N* but significantly decreases for $B≥3$ with $A>2$ (T-test, $p<0.05$). The exception again being at $K=0$ when there is no significant difference between $B=1$ and $B=2$ (T-test, $p<0.05$).

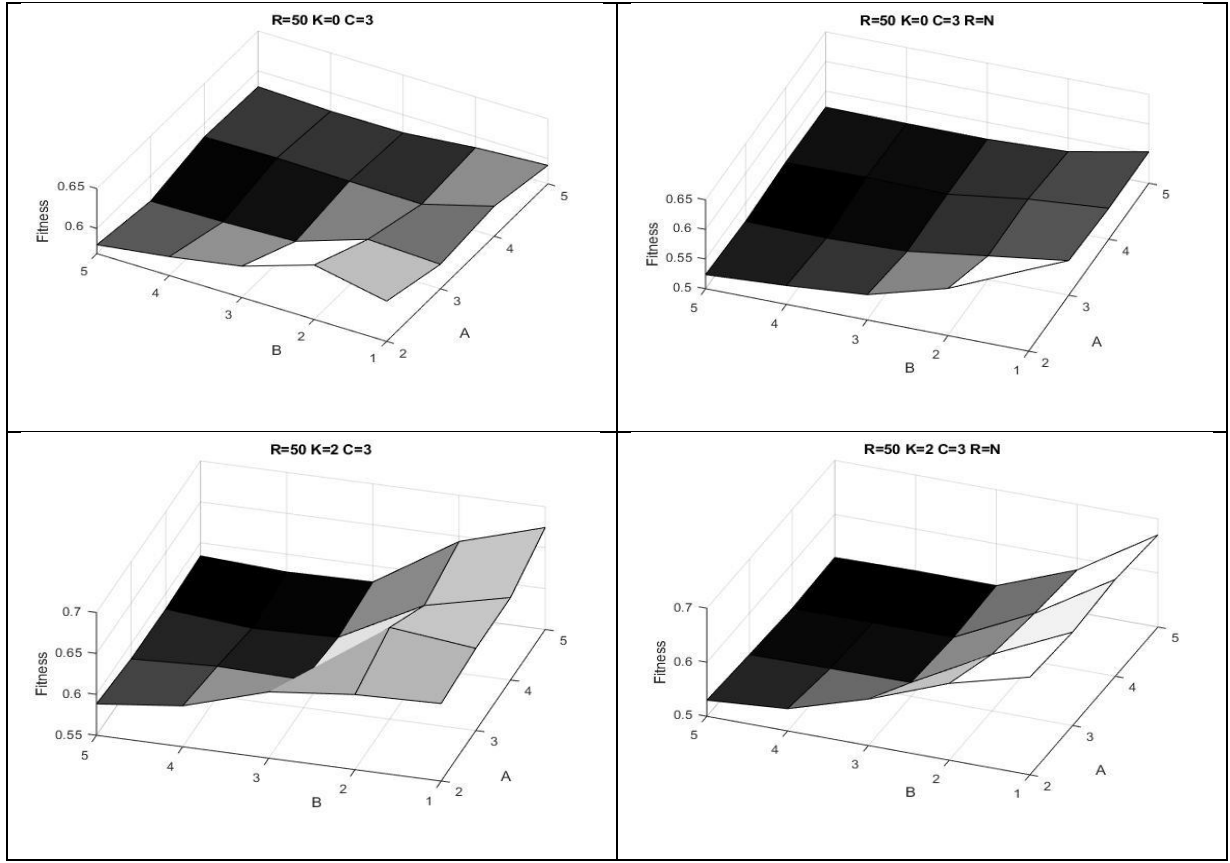

Figure 9. Showing the typical fitness reached by one species after 5000 generations with varying internal connectivity ($B$), number of explicit fitness traits ($N$), for various $K$ and $A$.

In the above, fitness is calculated from the state of the $N$ trait nodes on the step after $U$ network update cycles, i.e., typically within an attractor. To explicitly consider the evolution of temporal behaviour, i.e., particular sequences of gene activity, the state of the RMN can be sampled on every update cycle, i.e., up to and including within an attractor. Here total fitness is calculated as the average of the fitness of each successive state of the $N$ nodes for $U$ cycles. Thus, networks must evolve temporal behaviour which keeps them consistently within the high optima region(s) of the fitness landscape.

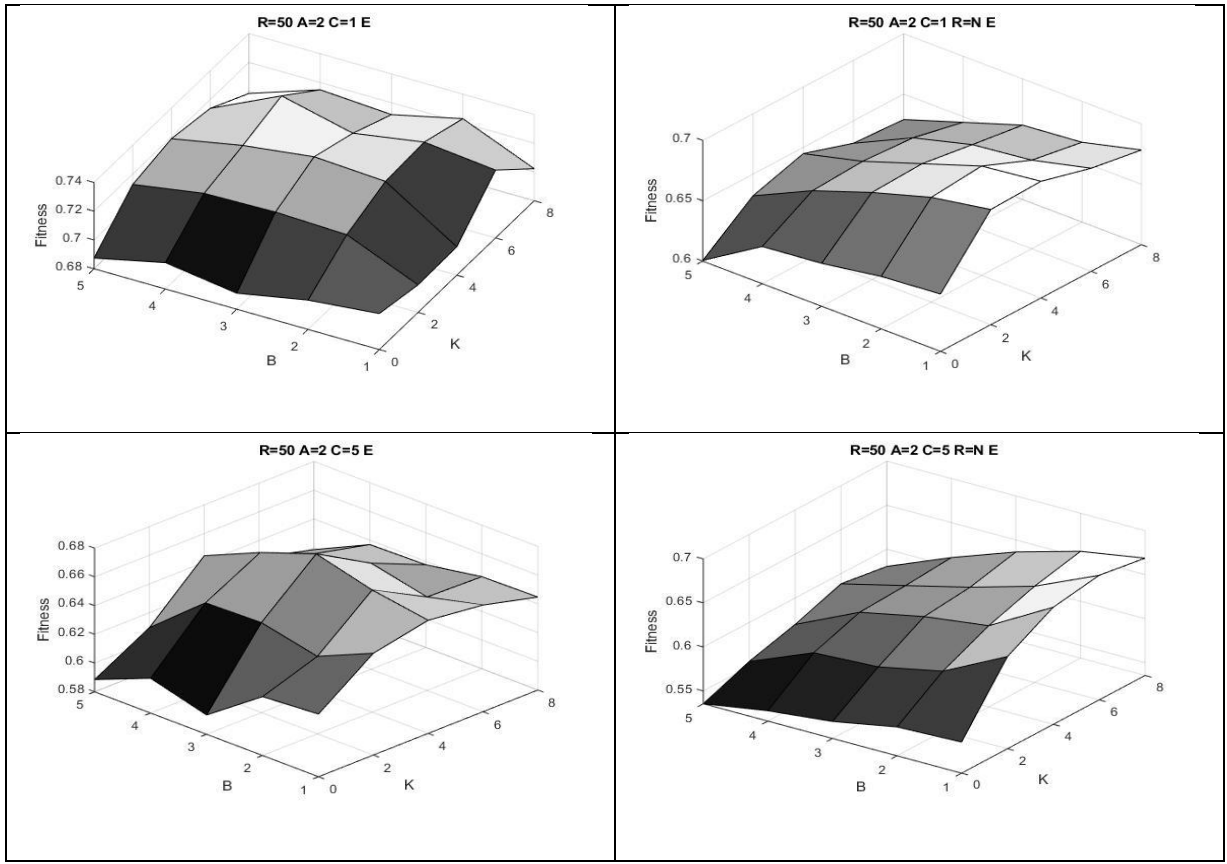

Figure 10. Showing the fitness reached by one species after 5000 generations on landscapes of varying ruggedness (*K*), coupling (*C*), number of explicit fitness traits (*N*), with *A*=2, and fitness calculated on each update step.

Figure 10 shows examples of how the change causes a significant increase in fitness (T-test, *p*<0.05) achieved with any *K* and *N*, with low *C*, for *B*>2 with *A*=2. Fitnesses are not significantly affected otherwise (T-test, p ≥0.05). Figure 11 shows examples of how the benefit for high *B* is lost when *A*>2, again showing a sharp decrease in relative fitness for *B*>2, as in Figure 9.

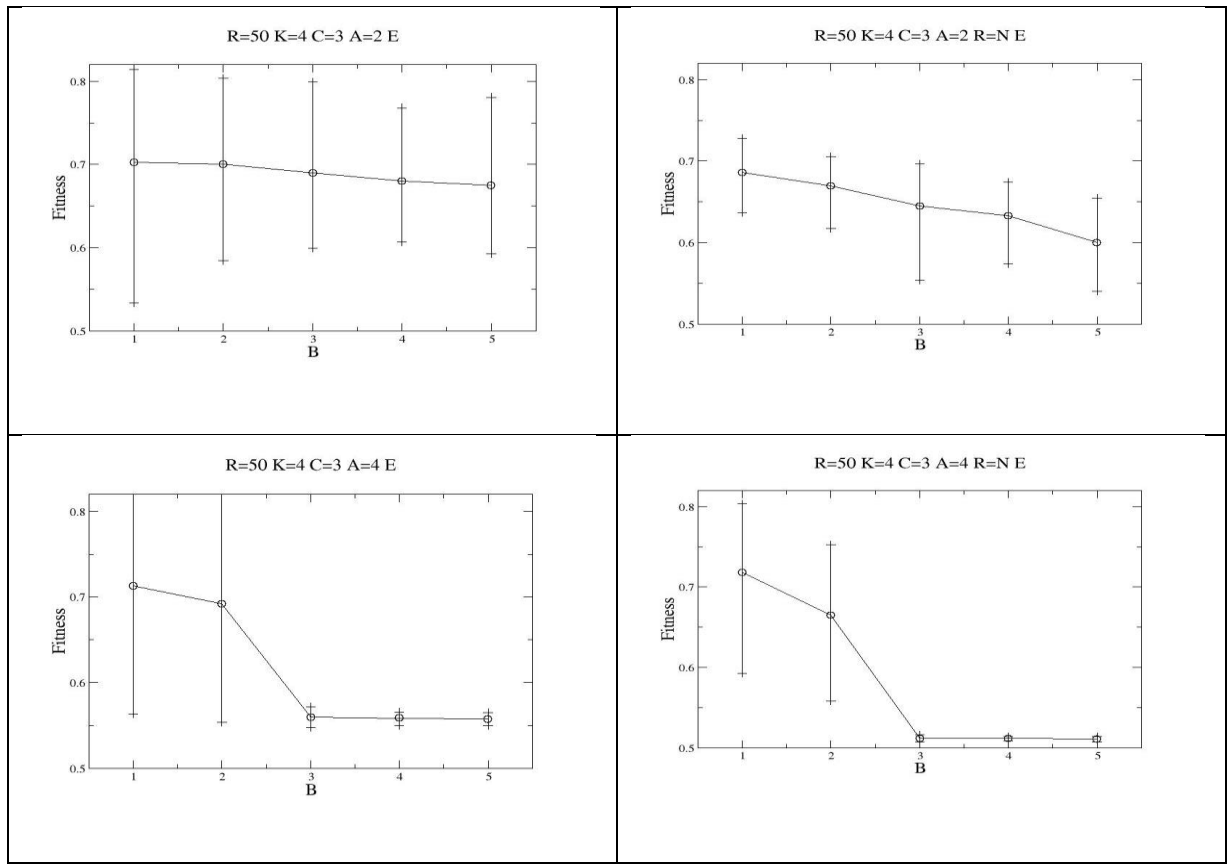

Figure 11. Showing the typical fitness reached by one species after 5000 generations with varying internal connectivity (*B*), number of explicit fitness traits (*N*), for *A*=2 compared to *A*=4, with *K*=4 and *C*=3, where the fitness is calculated on each update step.

Traditionally, RBN update synchronously, i.e., a global clock signal is assumed to exist but it has long been suggested that this assumption is less than realistic for natural systems. Harvey and Bossomaier [1997] were first to present an asynchronous form of RBN wherein a node is picked at random (with replacement) to be updated, with the process repeated *R* times per cycle to give equivalence to the synchronous case. The resulting loss of determinism means such networks no longer fall into regular cyclic attractors, rather they either fall into point attractors or so-called "loose" attractors containing a subset of possible states. Many forms of asynchronous updating are possible (e.g., see [Gershenson, 2004] for an overview) but the simple random scheme is used here to explore such updating in RMN.

Simulated evolution has previously been used with asynchronous RBN, beginning with attractor matching to exhibit defined rhythmic behaviour [DiPaolo, 2001].

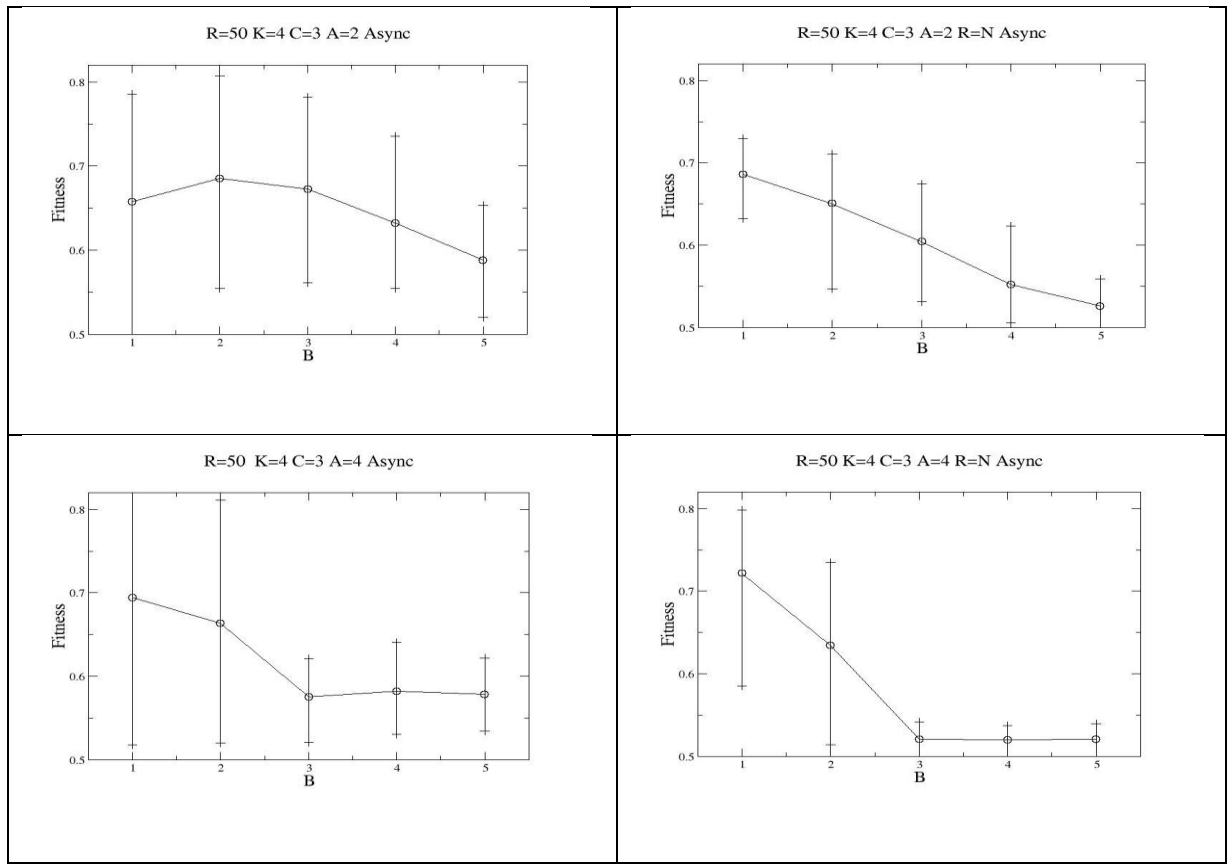

Figure 12. Showing the typical fitness reached by one species after 5000 generations with varying internal connectivity (*B*), number of explicit fitness traits (*N*), for *A*=2 compared to *A*=4, with *K*=4 and *C*=3, with asynchronous updating.

Figure 12 shows examples of how general behaviour remains largely unchanged in all cases with the asynchronous updating, although fitness is typically decreased for high *B* (>3) and *A*=2.

## 5. Asymmetry in the RMNKCS Model

Kauffman and Johnsen [1992] explored the effects of asymmetric *K* for varying *C* in the NKCS model. In particular, they describe the changes in behaviour seen before Nash equilibria are reached. That is, in the symmetrical case, as *C* increases, the time taken to reach a Nash equilibrium increases (see Figure 4) and, for a given *C*, the time taken to reach an equilibrium decreases with increasing *K*. They show how, for higher values of *C*, a low *K* species has a higher pre-Nash average fitness against a high *K* species, and vice versa. Other aspects of the NKCS model have recently been varied to create asymmetric pairings in other ways, including relative rates of evolution, degree of coupling, and rate of mutation [Bull, 2021]. Altering the degree of internal connectivity (*B*) and size (*R*) becomes possible in this model.

With low fitness landscape coupling, the internal connectivity of the partner (*B'*) has little affect upon the other, regardless of *A* (not shown). That is, fitness for a given *B* does not change significantly for any *B'*. As Figure 13 shows, with higher coupling (*C*>3), low values of internal connectivity appear mutually advantageous. That is, with *A*=2, both partners benefit if *B*<4 and *B'*<4. This region of beneficial connectivity decreases with *A*>2 to lower values of *B* and *B'*. The same general behaviour is seen for *R*=*N* and the fitness calculated on each step (not shown).

Figure 14 shows examples of the effects of varying the network size of the partner (*R''*). As can be seen, as the degree of coupling increases, for low internal connectivity, fitness increases/decreases significantly with a larger/smaller partner. It is well-established for *A*=2 that the typical number and size of attractors increases with *R* (e.g., see [Kauffmann, 1993]). The results here suggest this remains the case for *A*>2 and that a peak(s) in a fitness landscape can be more efficiently located under such conditions.

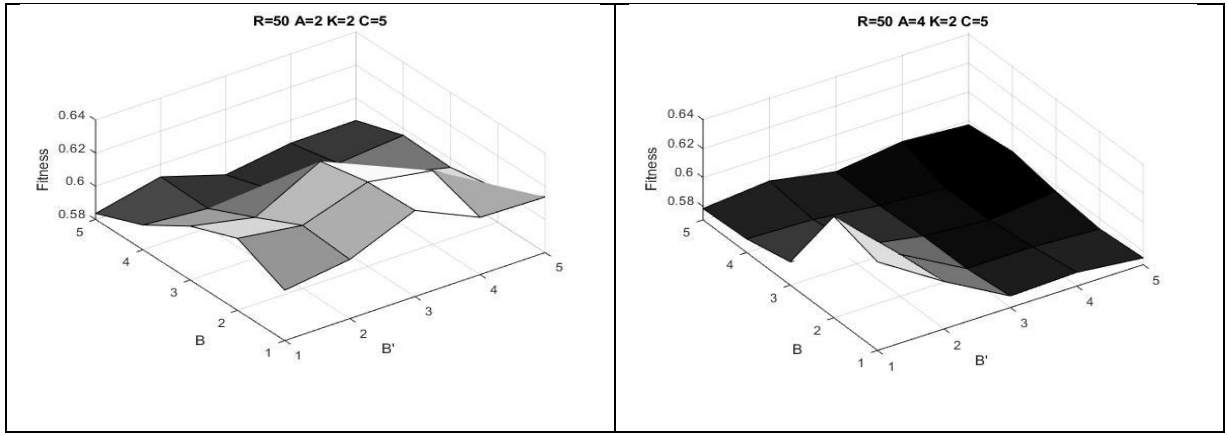

Figure 13. Showing the fitness reached after 5000 generations when the two partners have different degrees of internal regulatory network connectivity (*B*), for a high degree of landscape coupling (*C*), with *A*=2 and *A*=4 .

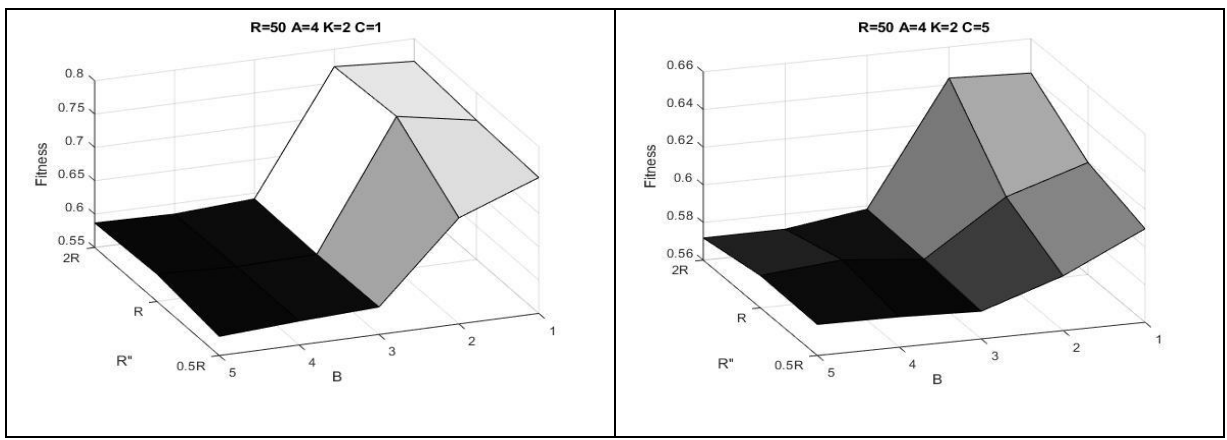

Figure 14. Showing the fitness reached after 5000 generations when the two partners have different sized regulatory networks (*R*), for varying internal network connectivity (*B*) and landscape coupling (*C*), with *A*=4 .

## 6. Conclusion

This paper has explored the coevolutionary behaviour of Boolean and multi-valued gene regulatory networks within tuneable fitness landscapes. Results indicate that low gene connectivity near or at the theoretical critical regime enables the highest fitness levels. That is, fitness levels are maximized on all landscape types used and this holds whether a constant or epochal state sampling scheme is adopted. The same results are true if asynchronously updating networks are used.

Boolean models have been used to accurately predict aspects of the regulatory dynamics seen in mammalian cells [Faur´e et al., 2006], Drosophila [Albert & Othmer, 2003], yeast [Li et al., 2004], amongst others. Despite this, their discretization of protein/RNA concentrations, failure to capture noise [Tkaˇcik et al., 2008], the possible existence of more than two levels of gene expression [Setty et al., 2003], etc. has motivated the development of multi-valued models. As noted above, formal analysis of increasing the number of gene states $A$ suggests the critical regime of connectivity tends towards $B$=1 [Solé et al., 2000]. However, biological data suggests 1.5 ≤ $B$ ≤ 2 [Leclerc, 2008], i.e., as somewhat predicted by the Boolean case. Such analysis assumes all multi-valued states and logic functions are equally likely which is perhaps unrealistic [Wittmann et al., 2010]. With simulated evolution able to shape the logic functions (and node connections) here, the fitness difference between $B$=1 and $B$=2 is typically not significant for $A$>2. The difference is significant when $R$=$N$ but that is often the case when $A$=2 as well (see Figures 8 and 9). The exception to the rule is when $K$=0 wherein there is no significant difference between $B$=1 and $B$=2 with $R$=$N$ and $A$>2 (Figure 9).

In the $R$=$N$ case each gene makes a direct contribution to fitness which is intended to be analogous to bacteria. Bacteria may be seen to evolve on fitness landscapes of relatively low ruggedness, i.e., $K$≈0. Therefore the results here support the biological data where 1.5 ≤ $B$ ≤ 2 since the conditions under which $B$=1 is superior to $B$=2 do not typically occur in nature: in bacteria, $R$=$N$ and $K$≈0; and, in eukaryotes, $R$>$N$ and $K$>0.